\documentclass[aps,prb,reprint,superscriptaddress]{revtex4-1}
\usepackage[latin1]{inputenc}	
\usepackage[T1]{fontenc}
\usepackage[cyr]{aeguill}
\usepackage[pdftex]{graphicx}
\usepackage{wrapfig}
\usepackage[english,french]{babel}
\usepackage{amsmath}
\usepackage{amssymb,amsfonts,textcomp}
\usepackage{color}
\usepackage{array}
\usepackage{hhline}
\usepackage{hyperref}
\usepackage{fancyhdr}
\usepackage[pdftex]{graphicx}
\definecolor{violet}{rgb}{0.5,0,0.5}
\definecolor{vert}{rgb}{0,0.65,0}
\usepackage{ulem}

\date{2011-4-5}

\begin{document}
\title{First principles study of iron-based molecule grafted on graphene}
\author{A. Reserbat-Plantey}

\affiliation{Institut N\'eel, CNRS-UJF-INP, BP 166, 38042 Grenoble cedex 9, France}
\author{P. Gava}
\affiliation{Institut de Min\'eralogie et Physique des Milieux Condens\'ees, CNRS-UMR 7590, Universit\'e Pierre et Marie Curie-Paris 6,
Universit\'e Denis Diderot-Paris 7, IPGP, F-75252 Paris, France}
\author{N. Bendiab}
\email{nedjma.bendiab@grenoble.cnrs.fr}
\affiliation{Institut N\'eel, CNRS-UJF-INP, BP 166, 38042 Grenoble cedex 9, France}
\author{A. M. Saitta}
\affiliation{Institut de Min\'eralogie et Physique des Milieux Condens\'ees, CNRS-UMR 7590, Universit\'e Pierre et Marie Curie-Paris 6,
Universit\'e Denis Diderot-Paris 7, IPGP, F-75252 Paris, France}
 \begin {abstract}{Motivated by recent experimental studies on single molecular magnets grafted on graphene and single walled carbon nanotubes, we investigate the structural, electronic, and magnetic properties of an iron based magnetic molecule grafted on a graphene sheet using \textit{ab initio} calculations. 
 In particular, the induced charge transfer and magnetization are described in terms of the coupling between the molecule and the graphene orbitals. 
 This interaction and its effects on graphene electronic properties are determined and discussed in view of the potential utilization of graphene in spintronics.}
\end{abstract}
\maketitle
\clearpage

%%%%%%%%%%%%%%%%%%%%%%%%%%%%%%%%%%%%%%%%%%%%%%%%%%%%%%%%%%%%%%%
%%%%%%%%%%%%%%%%%%%%%%%%%INTRODUCTION%%%%%%%%%%%%%%%%%%%%%%%%%%%%%%
%%%%%%%%%%%%%%%%%%%%%%%%%%%%%%%%%%%%%%%%%%%%%%%%%%%%%%%%%%%%%%%

\section{Introduction}
Graphene-based systems have recently attracted much interest from both experimental and theoretical aspects.
Its chemical inertness, hydrophobic behavior, large electron mobility, scalable production and intrinsically low spin-orbit coupling makes it a very promising candidate for sensors\cite{schedin}, electronics \cite{Neto}, spintronics \cite{{tombros},{Candini2011}} and nanomechanics \cite{{Bunch},{Bachtold}}. 
Experimentally, the graphene capability to detect small charge transfer effects by grafting molecules\cite{{schedin},{haddon}}, but also to induce superconductivity\cite{Kessler,Heersche}, as well as Kondo effect\cite{Krasheninnikov} by deposing metals have been investigated. 
 More recently, surface-enhanced Raman signal has been realized on pyrene-based molecules grafted on graphene \cite{{Dress}, {Bendiab}}, allowing a detection down to the limit of few isolated molecules.
  Detection of single molecule magnet\cite{{Bendiab},{matias},{WWreview}} by using graphene or carbon nanotubes could also be a way to probe magnetic properties at the single molecule level.
Theoretically, \textit{ab initio} calculations on magnetic properties of graphene based materials have been performed\cite{Xiao,Leenaerts}. 
In particular, \textit{ab initio} studies of adatoms on graphene \cite {Johl,Mao,Nieminen2003} magnetic Co dimers\cite{Xiao} and even small magnetic molecules such as O$_2$, NO or NO$_2$\cite{Leenaerts} have been reported.

The interaction between the graphene monolayer and a magnetic molecule is an important point in order to have insights on local effects such as induced magnetization and charge transfer, which are of interest for spintronic and nano-electronic applications.
In the present work, we chose the iron tetraphtalic acid molecule - $FeTPA_{4}$ \cite{Gambardella} - deposited on monolayer graphene as a case study of this interaction. 
Indeed, Gambardella et al. \cite{Gambardella} have succeeded in the manipulation of the magnetic anisotropy of a supramolecular assembly of $FeTPA_{4}$ self assembled on a Cu surface. 
However, to speed up our DFT calculations, we modeled $FeTPA_{4}$ by explicitly studying its magnetic core  $Fe(OH)_{4}$, as we will discuss in the following, where this choice will be justified on the basis of structural, electronic, and magnetic properties.

%%%%%%%%%%%%%%%%%%%%%%%%%%%%%%%%%%%%%%%%%%%%%%%%%%%%%%%%%%%%%%%
%%%%%%%%%%%%%%%%COMPUTATIONAL DETAILS%%%%%%%%%%%%%%%%%%
%%%%%%%%%%%%%%%%%%%%%%%%%%%%%%%%%%%%%%%%%%%%%%%%%%%%%%%%%%%%%%%

\section{Computational details}

First principles DFT calculations were performed within the plane wave approximation (PW) and pseudopotential scheme, as implemented in the quantum ESPRESSO code \cite{QE, PW}.
We adopt a Perdew-Burke-Ernzerhof (PBE) \cite{Perdew} gradient corrected functional for the exchange and correlation potential, and ultrasoft pseudopotential technique is used to describe C, H, O and Fe atoms \cite{LS}. 
Electronic wave functions are written in terms of plane waves, with an energy up to 30 Ry, which is sufficient to ensure convergence of structural, electronic and magnetic properties. 
The electronic occupation is computed using Fermi-Dirac distribution, with a smearing parameter of 136 meV, which corresponds to an electronic temperature of 1578 K.
The Brillouin zone integration is performed with a uniform \textbf{k} points grid of (15x15x1).
The convergence of the electronic and spin properties with the electronic temperature and k-points grid has been tested.
 Our choice of 1578 K with  (15x15x1) k-points grid is within the range of typical electronic temperatures used to describe graphitic systems, and it is motivated by the fact that such values ensure converged results with reasonable computational effort.
Structures are relaxed until forces are below 0.02 eV/\r{A}.
We used a supercell composed by (6x6) repetition of a graphene unitary cell in order to ensure a negligible interaction between
periodic images of the adsorbed molecule.
Spin polarized calculations are also performed, and we consider polarization along the z axis (\textit{i.e.} perpendicular to graphene plane). \\

\begin{figure}[htbp]
\begin{center}
       \includegraphics[width=9cm]{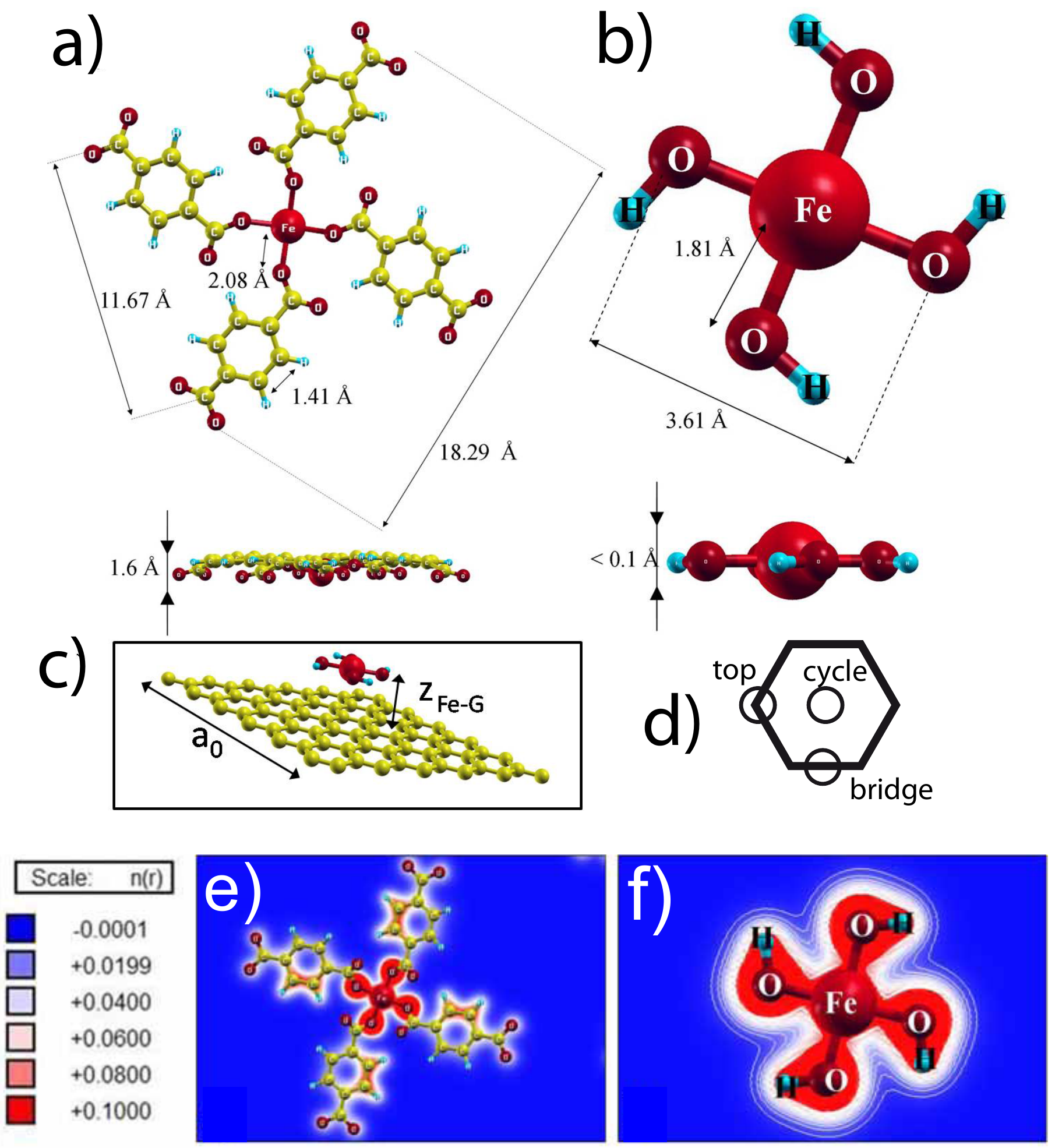}
         \caption{(Top and middle panels) Structural description of: a) $Fe(TPA)_4$ isolated molecule, b)
         $Fe(OH)_4$ isolated molecule, c) hybrid system composed of graphene and $Fe(OH)_4$. Iron and oxygen atoms are indicated in red,
         hydrogen atoms in blue, and carbon atoms in yellow. The high symmetry adsorption sites are indicated with respect
         to the hexagonal cell of graphene, (d). Bottom panel: charge density map within xy plan for e)
         $Fe(TPA)_4$ and f) $Fe(OH)_4$.}
\label{fig:Figure1}
\end{center}
\end{figure}

\subsection{Modeling $FeTPA_{4}$ through a smaller molecule}
The $FeTPA_{4}$ molecule is difficult to simulate due to its large number of atoms (33 atoms) and, especially, of the size of the graphene supercell that should contain it.
As mentioned above, however, the $Fe(OH)_4$ molecule mimics well the $Fe(TPA)_4$ molecule. 
This smaller, case-study system contains the same $FeO_4$ core as the bigger one, but the four large pyrenic arms are simply replaced by hydrogen atoms.
As shown in the top panels of Fig. \ref{fig:Figure1}, the two molecules have a similar planar structure.
The charge distribution around the iron core are computed projecting all the electronic wave functions $\varphi_{i,\sigma}$ on the atomic orbitals centered in each given atom through the equation:

\begin{equation}
n^{\beta} = \sum _{i,   \ell, m,\sigma=\uparrow,\downarrow} \left|\left\langle \phi_{\ell,m}^{\beta} | \varphi_{i,\sigma} \right\rangle \right|^{2},
\label{eq:N_projection}
\end{equation}

where $\phi_{n \ell}^{\beta}$ is the atomic-like orbital for atom $\beta$, labelled with its energy and angular momentum quantum number $m,l$.

\begin{table}[htbp]
\begin{center}
\begin{tabular*}{8cm}{@{\extracolsep{\fill}}clcccc}
\hline
\hline

Molecule    & Atom  & \multicolumn{2}{c}{Lowdin charge}         & Polarization \\

\hline
\hline

$Fe(OH)_{4}$        &&      &           &\\
        &Fe     &15.628         &           &3.196 \\
        &       &\textit{s}     &2.503      &\\
        &       &\textit{p}     &6.722      &\\
        &       &\textit{d}     &6.402      &\\
        &O      &6.450          &           &0.186\\
        &       &\textit{s}     &1.686  &\\
        &       &\textit{p}     &4.764      &\\

 \hline

$Fe(TPA)_{4}$       &&      &           &\\
&Fe     &15.611         &           &3.102\\
        &       &\textit{s}     &2.473              &\\
        &       &\textit{p}     &6.741              &\\
        &       &\textit{d}     &6.395              &\\

        &O*         &6.246          &                   &0.122\\
        &       &\textit{s}     &1.623              &\\
        &       &\textit{p}     &4.623              &\\

\hline
\hline

\end{tabular*}

\caption{Calculated Lowdin charges (see eq.\ref{eq:N_projection}) and their
decomposition onto atomic-like orbitals, for
         Fe and O atoms in $Fe(OH)_4$ and $Fe(TPA)_4$ molecules.  O* stands for first neighbors of Fe atom.
         Polarization is computed along the z direction for Fe and O atoms as the difference between spin up and down charges.}
\label{charge_proj}
\end{center}
\end{table}

We observe in Table \ref{charge_proj} that the charge associated to the core iron atom (Lowdin charge) remains very similar in both cases.
The only slight charge difference between the two cases concerns the O atoms in $Fe(OH)_4$ having around 0.2 electrons more
than in $Fe(TPA)_4$.
We also computed the spin polarization along the $z$ direction for Fe and O atoms, as the difference between the spin up and down charges.
As shown in Table \ref{charge_proj}, the spin polarization of Fe atoms is very similar, being 3.20 and 3.10 $\mu_B$ in $Fe(OH)_4$ and $Fe(TPA)_4$ molecules respectively.
In both cases this polarization comes from unpaired electron in 3d iron orbitals. 
The polarization of the O atoms, which comes from 2p orbitals, is slightly higher in the $Fe(OH)_4$ molecule.
The total polarization of the core is 3.94 and 3.59 $\mu_B$ in $Fe(OH)_4$ and $Fe(TPA)_4$, respectively, which corresponds to a variation of around 10$\%$. 
We can thus safely consider these two molecules equivalent from the spin magnetic point of view in the void; given the relative inertness of graphene sheets with respect to full metallic surfaces, we can reasonably extrapolate a very similar behavior when grafted on a graphene sheet.
\subsection{Calculation details of the hybrid system}
We use, to study the hybrid system composed by the $Fe(OH)_4$ molecule adsorbed on graphene, an hexagonal supercell composed of 36 unit cells (Fig.\ref{fig:Figure1}c) with a lattice parameter in the xy plane of a$_0$ = 14.76 \r{A}, and of 11.10 \r{A} in z direction. 
Due to Brillouin zone refolding, the Dirac (K) point of graphene refolds on the $\Gamma$ point. 
The three high symmetry adsorption sites of $Fe(OH)_4$ on graphene are shown in Fig.\ref{fig:Figure1}d.
We find that the most energetically favorable adsorption site is the top site, and only this one will then be considered in the following.
%%%%%%%%%%%%%%%%%%%%%%%%%%%%%%%%%%%%%%%%%%%%%%%%%%%%%%%%%%%%%%%
%%%%%%%%%%%%%%%RESULTATS ET DISCUSSION)%%%%%%%%%%%%%%%%%
%%%%%%%%%%%%%%%%%%%%%%%%%%%%%%%%%%%%%%%%%%%%%%%%%%%%%%%%%%%%%%%

\section{Results and discussion}

\subsection{Structural properties}

\begin{figure}[htbp]
\begin{center}
       \includegraphics[width=8cm]{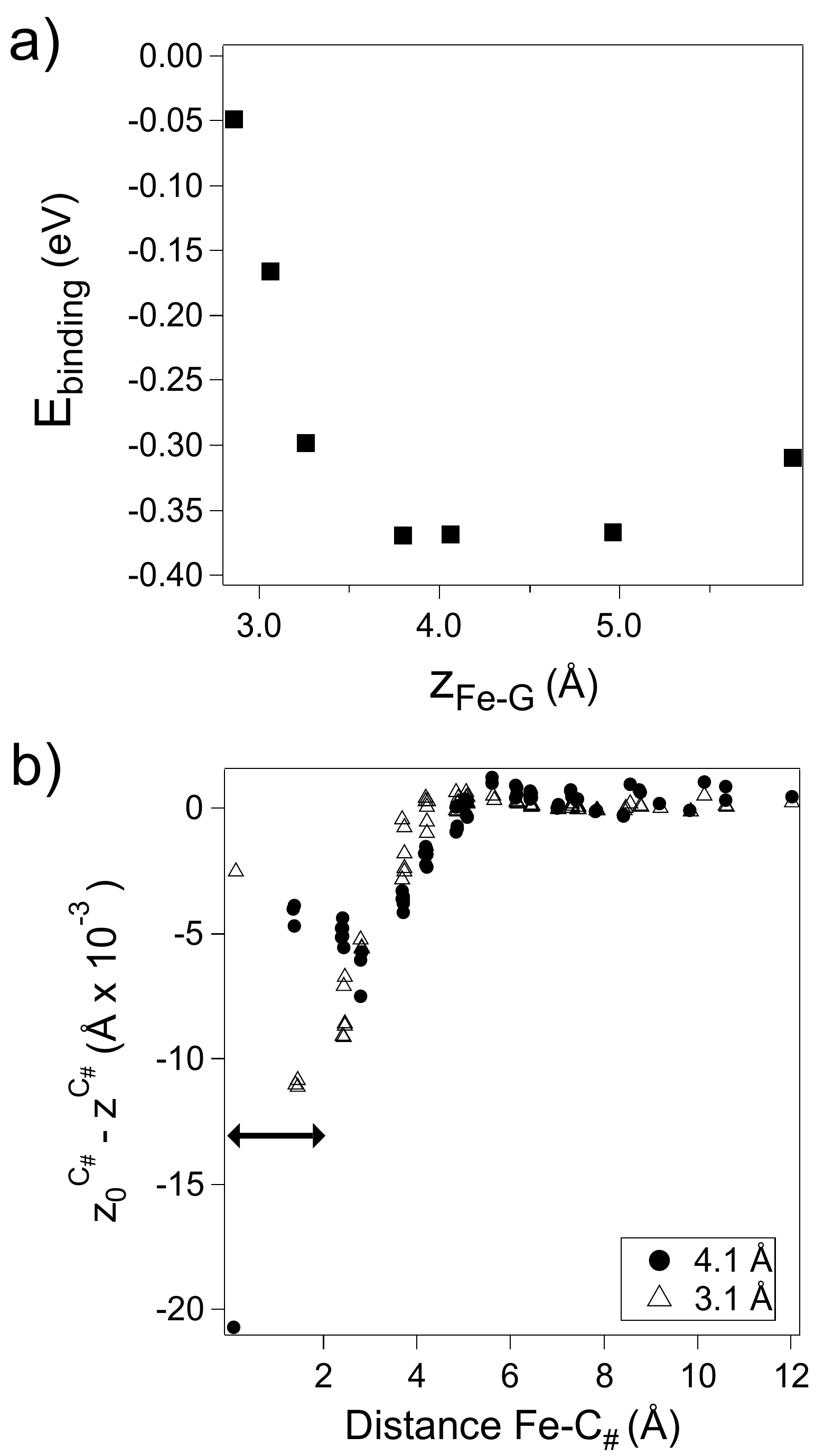}
         \caption{(a) Binding energy of the hybrid system as a function of the distance along z between the graphene surface and the iron atom
         of the Fe(OH)$_{4}$ molecule. (b) Displacements ($\Delta z^{C\#} $) of graphene C atoms positions along z with respect to the
         isolated case, as a function of distance, within the xy plane, to the adsorption (top site) C$_{top}$ atom. The
horizontal arrow represents the planar extension of $Fe(OH)_4$ molecule.
Two fixed values of z$_{Fe-G}$ are considered : the equilibrium distance
4.1 \r{A} (filled circles) and the characteristic distance 3.1 \r{A} (empty
triangles)}
         \label{fig:Figure2}
\end{center}
\end{figure}

We performed the structural relaxation of the hybrid system by setting a fixed distance z$_{Fe-G}$ between the Fe atom in the $FeOH_{4}$ molecule and the graphene plane. 
According to our calculations, top adsorption site is found to be the most stable from an energetic point of view and we defined z$_{Fe-G}$ as the distance along z between the Fe atom and the first C atom below, which we indicate as C$_{top}$. 
We calculated the binding energy as a function of this distance (see Fig. \ref{fig:Figure2}), as $E_{binding} = E_{graphene + FeOH_{4}} - E_{graphene} - E_{FeOH_{4}}$. 
The obtained equilibrium distance for the system is z$_{Fe-G}^{eq}$=4.1 \r{A}, and the corresponding binding energy is $E_{binding} = -0.36$ eV. 
However, it is well known that \textit{ab initio} DFT calculations do not take into account Van der Waals interactions, which are of course important in
weakly interacting systems. 
As a result, the distance we obtain is likely to be overestimated. 
A properly \textit{ab initio} correction should explicitly take into account these contributions in the DFT functional. 
However, this procedure is rather cumbersome and computationally costly, and a common approximation consists in carrying out the calculations at the experimental equilibrium distance. 
The most flagrant case is graphite, a 3D crystal formed by layers of graphene, whose experimental interplane equilibrium distance is about 3.35 \r{A}, while the DFT equilibrium value is more than 5 \r{A}. 
In an analogous way, in the following we always compare results obtained for the DFT equilibrium distance z$_{Fe-G}^{eq}$=4.1 \r{A} with results obtained at a shorter distance, z$_{Fe-G}$= 3.1 \r{A} which is comparable to the characteristic interaction distance observed in sp$^{2}$ carbon materials \cite{cardona,dresselhaus}.\\
In both cases, the flatness of the graphene plane is slightly perturbed by the presence of the molecule.
While C-C bond distances are practically unaffected ($\leq$ 0.1 $\%$), the z coordinate of C atoms in the relaxed hybrid system changes by as much as 1 $\%$, as shown in Fig. \ref{fig:Figure1}.
This surface modulation extends up to a distance of around 5.5 \r{A} whether z$_{Fe-G}$ equals 4.1 or 3.1 \r{A}.

\subsection{Electronic properties}

The charge transfer between graphene layer and the molecule can be quantified by computing the difference between the total charge of the graphene sheet in the hybrid system and of isolated graphene.
We obtain a charge difference of -0.63 electrons, revealing an induced hole doping on graphene.
This charge transfer is equal to a surface charge density of about -3.4 10$^{13}$ cm$^{-2}$, which is coherent with common predictions and experiments involving graphene \cite{Novoselov2, Novoselov3, Ferrari}.\\
We also evaluate the local electronic charge transfer for each C atoms in the graphene plane and extract contributions from the different orbitals by projecting the electronic eigenstates on the atomic wavefunctions as in Eq. \ref{eq:N_projection}. 
At a molecule-graphene distance of 4.1 \r{A} we observe that the charge transfer from the graphene sheet originates mostly (98\%) from the 2p orbitals of C atoms. 
A maximum charge transfer is observed for 2p states of C atoms at 2.5 \r{A} from the Fe center, while at distances greater than 6 \r{A} the spatial charge transfer remains constant. 
For $z_{Fe-G} = 3.1 $\r{A}, a similar behavior is observed.
From this result we determine an interaction length of about 6 \r{A}, which is in agreement with the one estimated by the analysis of the surface corrugation (see Fig.\ref{fig:Figure1}).
Moreover, the comparison between the electronic band structure of isolated graphene and of the hybrid system (not reported) indicates that, despite a shift of the Fermi level, the perturbation induced by the molecule is negligible, as also shown by the non-dispersive character of the molecular states. \\
This analysis of the charge transfer and of the electronic band structure of the hybrid system suggests  that the quasi-metallic character of graphene is not perturbed by its interaction with the grafted magnetic molecule.
This result is promising in view of the use of graphene as a sensitive detector.
\subsection{Magnetic properties}

As mentioned above, we also performed spin polarized calculations on isolated molecule and on the hybrid system. 
The obtained magnetic moment for the isolated molecule $Fe(OH)_{4}$ is about 3.94 $\mu_B$, whereas in the hybrid system it is about 3.34 $\mu_B$. 
The magnetization difference between the two systems is about 0.6 $\mu_B$ indicating that the interaction with the graphene substrate generates 0.6 ``paired'' electrons in this hybrid system which is coherent with the estimated charge transfer of 0.63 electrons found in the previous section. 
We then analyze this variation of the molecular magnetization within the hybrid system spin states occupations. 
Thus, we calculated the projected density of states (PDOS) for spin up and spin down states within the hybrid system. 
More precisely, we focused on the projection into 2p$_z$ states of carbon atoms, which are the most sensitive to their environment.\\

\begin{figure}[htbp]
\begin{center}
      \includegraphics[width=8cm]{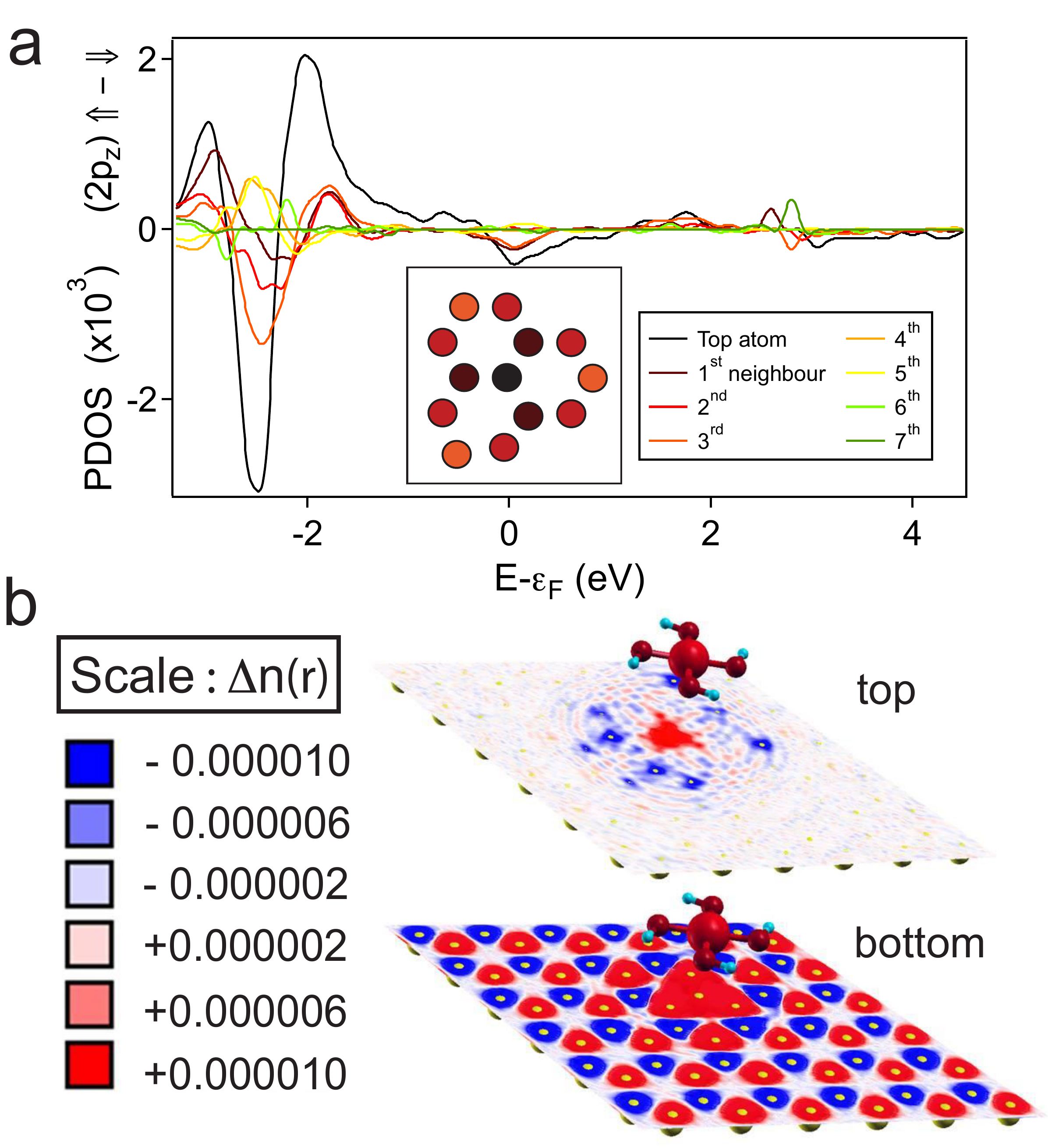}
         \caption{(a) Difference between PDOS (see text) for spin up and down states within the hybrid system.
Projections are performed onto carbon 2$p_z$ atomic orbitals. Carbon atoms
are labelled by their neighbouring rank, in the xy plane, with respect to C$_{top}$. 
The sketch represents the C atoms of the graphene. $Fe(OH)_4$ is located at 3.1 \r{A} vertically above the black C$_{top}$ atom and the different colors indicates the different neighboring rank.
$PDOS_{2p_z}(\uparrow) - PDOS_{2p_z}(\downarrow)$ is averaged for every
$C^{\#}$ atom at the same distance from C$_{top}$. (b) Spatial distribution
of the magnetization $\Delta n(\vec{r}) = n(\vec{r})_\uparrow -
n(\vec{r})_\downarrow$ within hybrid system for two distances z$_{Fe-G}$ =
4.1 \r{A} (top) and z$_{Fe-G}$ = 3.1 \r{A} (bottom).} \label{fig:Figure3}
\end{center}
\end{figure}

In Fig. \ref{fig:Figure3}a we report the difference between the PDOS for spin up and down states for different C atoms, labelled with their distance in the xy plane with respect to Fe atom in the $Fe(OH)_{4}$ molecule. 
This figure shows an energy range [-3 ; -1.5] eV where the PDOS for spin up and spin down differs and, in particular, for carbon atoms located up to the 5$^{th}$ neighbour (\textit{ie.} 5 \r{A} from C$_{top}$ atom).
This spin-interaction distance is consistent with the previously discussed structural and electronic interaction lengths of the hybrid system.\\
To go further, we analyze the mixing between atomic orbitals of $Fe(OH)_4$ molecule and the ones of graphene in the hybrid system. 
In order to quantify the hybridization level of the occupied states, we define the quantity $c_{g/M}$:

\begin{equation}
c_{g/M} = \sum_{AO} c^{AO}_{g,M}
\label{eq:coeff}
\end{equation}

where $C^{AO}_{g,M}$ represents the projection of an electronic states into atomic orbitals (AO) belonging to graphene atoms (g) or to the molecule atoms (M).
These states in the range [-3.2;-2.0] eV, with respect to the Fermi level, correspond to occupied electronic states where $c_{g}/c_{M}\in[10\% ; 90\%]$. 
Within that interval, both systems (\textit{i.e.} graphene and molecule) have a significant contribution to the global electronic wavefunctions which suggest a remarkable hybridization. 
This is consistent with findings shown in Fig. \ref{fig:Figure3}a, where we observe strongly polarized states in a region between [-3.0 ; -1.5] eV and also with the existence of additional mixed states .
In particular, at \textbf{k}=0 and at -2.4 eV there is an electronic state which is particularly important since it represents the overlap of Fe 3d$_{z^2}$ (13 $\%$) and top C atom 2p$_{z}$ orbital (8 $\%$). 
Moreover, the contribution of this atomic orbital from the closest carbon atom is not present in any other mixed wavefunction.
This indicates, for instance, the negligible interaction between the C atom and the oxygen atoms. All other Fe states are only present in pure molecular electronic states and not in mixed states.
Conversely, we do not observe strongly hybridized spin down states. 
The hybridized states within  [-3.2;-2.0] eV show mainly overlap between oxygen 2p$_{x,y}$ states and 2p$_{z}$ states of carbon atoms located at distances $\geq 5.5 \r{A}$ from Fe atom. 
On the basis of our findings, we expect a spin up charge excess around the carbon atoms closer to Fe atom and an excess of spin down charges within a crown of carbon atoms around 5.5 \r{A} from Fe atom. 
This local magnetization within graphene plane is shown in Fig. \ref{fig:Figure3}b. 
The induced magnetization in the graphene plane is suggested by concentric crowns of spin up/down electron in 2p states of carbon atoms.

\section{Conclusions}

We report a case study of an iron-based magnetic molecule grafted on a graphene sheet and, through \textit{ab initio} DFT calculations, we predict an induced magnetization effect to the graphene substrate.
We also describe a typical interaction length of 5-6 \r{A} within the graphene plane with respect to structural deformations, localized charge transfer and induced magnetization. 
The latter originates from a significant coupling between 2p$_z$ and Fe 3d$_{z^2}$  orbital for the closest carbon atom. 
On the other hand, we observe that the presence of an iron-based magnetic molecule in the close vicinity of a graphene sheet does not strongly perturb the electronic properties of the latter, thus providing a promising perspective to the design of non-destructive sensor devices.

We acknowledge useful discussions with X. Blase, V. Bouchiat, A. Candini, M. Urdampiletta and W. Wernsdorfer. 
Fig. 1 and 3 have been performed with XCrysden package \cite{xcry}. 
Calculations were done at IDRIS (Orsay, France), Project No. CP9-91387.

%%%%%%%%%%%%%%%%%%%%%%%%%%%%%%%%%%%%%%%%%%%%%%%%%%%%%%%%%%%%%%%%%%%%%%%%%%%%%%%%%%%%%%%%%%%%%%%%%%%%%%
%%%%    BIBLIOGRAPHY             %%%%%%%%%%%%%%%%%%%%%%%%%%%%%%%%%%%%%%%%%%%%%%%%%%%%%%
%%%%%%%%%%%%%%%%%%%%%%%%%%%%%%%%%%%%%%%%%%%%%%%%%%%%%%%%%%%%%%%%%%%%%%%%%%%%%%%%%%%%%


\begin{thebibliography}{breitestes Label}
\bibitem{LS}
Spin orbit coupling is not considered in this study.
\bibitem{note}{Note that the \textbf{K} point in the first Brillouin zone corresponds to \textbf{k}=0 since we are using a super-cell which is 6x6 graphene unit cell.}
\end{thebibliography}
\end{document}